\documentclass[ACS]{WileyNJD-v1}

\articletype{Article Type}%

\received{00 XXX 0000}
\revised{00 XXX 0000}
\accepted{00 XXX 0000}

\begin{document}

\title{Fluid and kinetic modelling for non-local heat transport in magnetic fusion devices}

\author[1]{Guido Ciraolo*}

\author[1]{Hugo Bufferand}

\author[2,3]{Pierfrancesco Di Cintio}

\author[1]{Philippe Ghendrih}

\author[4,3]{Stefano Lepri}

\author[5,3,4]{Roberto Livi}

\author[6]{Yannick Marandet}

\author[7]{Eric Serre}

\author[1]{Patrick Tamain}

\author[1]{Matteo Valentinuzzi}

\authormark{Guido Ciraolo \textsc{et al.}}

\address[1]{\orgdiv{IRFM}, \orgname{CEA}, \orgaddress{\state{F-13108 St Paul Lez Durance}, \country{France}}}

\address[2]{\orgdiv{IFAC}, \orgname{CNR}, \orgaddress{\state{I-50019 Sesto Fiorentino}, \country{Italy}}}

\address[3]{\orgdiv{INFN}, \orgname{Sezione Firenze}, \orgaddress{\state{I-50019 Sesto Fiorentino}, \country{Italy}}}

\address[4]{\orgdiv{ISC}, \orgname{CNR}, \orgaddress{\state{I-50019 Sesto Fiorentino}, \country{Italy}}}

\address[5]{\orgdiv{Dipartimento di Fisica e Astronomia and CSDC}, \orgname{Universitá di Firenze}, \orgaddress{\state{I-50019 Sesto Fiorentino}, \country{Italy}}}

\address[6]{\orgdiv{PIIM}, \orgname{CNRS - Aix-Marseille Université}, \orgaddress{\state{F-13397 Marseille}, \country{France}}}

\address[7]{\orgdiv{M2P2}, \orgname{CNRS - Aix-Marseille Université}, \orgaddress{\state{F-13451 Marseille}, \country{France}}}

\corres{*Guido Ciraolo, CEA IRFM, 13108 St Paul Lez Durance. \email{guido.ciraolo@cea.fr}}

\abstract[Summary]{In order to improve the presently used ad hoc flux limiter treatment of parallel heat flux transport in edge plasma fluid codes we consider here a generalized version of the Fourier law implementing a non-local kernel for the heat flux computation. The Bohm boundary condition at the wall is recovered introducing a volumetric loss term representing the contribution of suprathermal particles to the energy out flux. As expected, this contribution is negligible in the strongly collisional regime while it becomes more and more dominant for marginally and low collisional regimes. In the second part of the paper, we consider a kinetic approach where collisions are considered using the Multi-Particle-Collision (MPC) algorithm. Kinetic simulation results at medium and low collisionality are also reported.}
\maketitle
\section{Introduction}\label{intro}
\begin{figure}
\begin{center}
\includegraphics[width=0.9\textwidth]{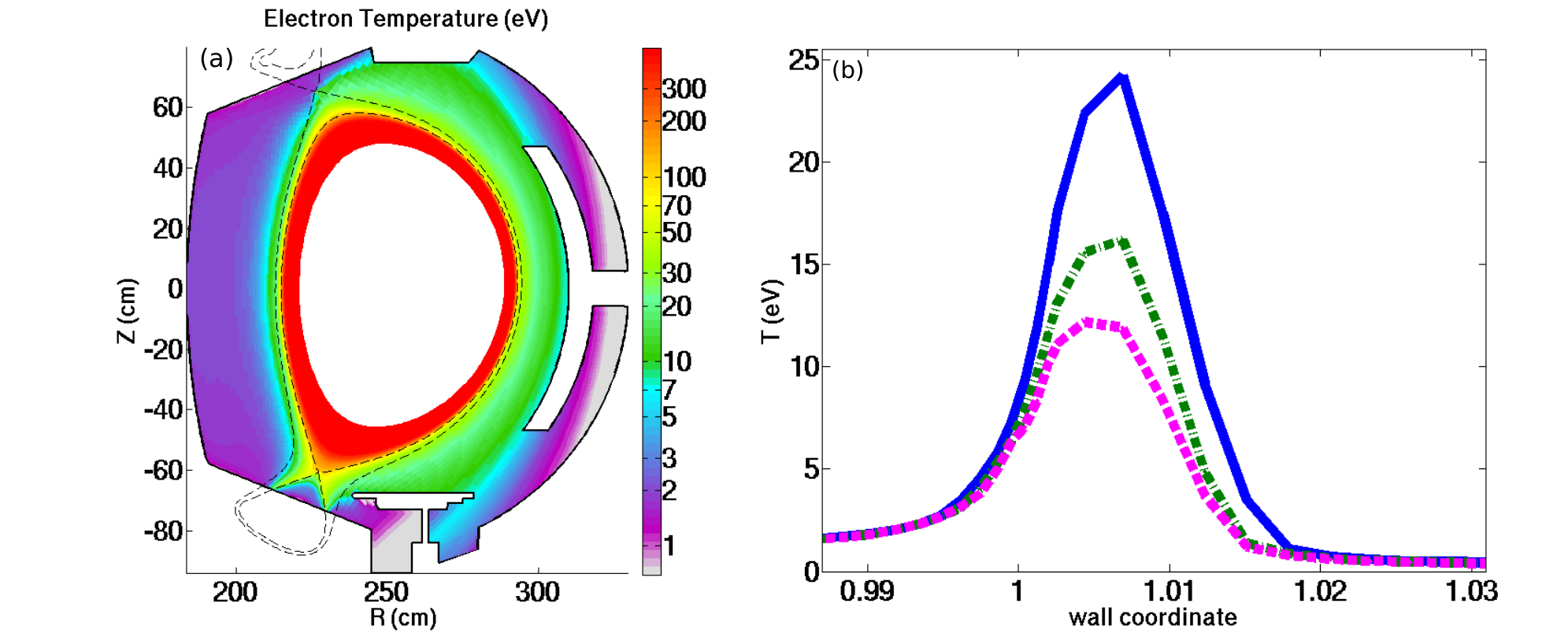}
\end{center}
\caption{Contour plot of the electron temperature in the poloidal section obtained from a SOLEDGE2D-EIRENE simulation (using the SH expression for the electron heat flux computation) with input power $P_{in}=4MW$ and a gas puff of $4\times 10^{21}$ atoms per second activated in the private flux region (a). Electron temperature profiles on the outer divertor target computed using SH expression (solid blue line), flux limiter expressions with parameter $\alpha=0.3$ (dash-dot green line), and $\alpha=0.15$ (dashed pink line) (b).}
\label{Soledge_2D_temp}
\end{figure}

Modelling parallel heat transport in edge tokamak plasma is a crucial issue for predictions of power loads on divertor targets. In the operational regimes of interest for a magnetic fusion device a significant temperature gradient will build up along the field line between the upstream hot region that acts as a heat source, and the colder plasma region at the wall that acts as a sink. Numerical estimations of edge and SOL plasma rely mainly on 2D transport codes like e.g. SOLEDGE2D \cite{HB2015}, SOLPS-ITER\cite{SW2015}, EDGE2D \cite{CG2014}, SONIC\cite{shimizu09}, UEDGE\cite{uedge94}. These numerical tools are based on a fluid approach and a collisional closure with the so-called Spitzer-H{\"a}rm (hereafter SH, see Ref. \cite{sh}) expression for the parallel heat flux
\begin{equation}
q_{\parallel}(x)=-\kappa(x)\nabla_{\parallel}T(x),
\end{equation} 
where the thermal conductivity $\kappa(x)$ is computed in the strong collisionality assumption (i.e. considering a small departure from the Maxwellian distribution function), and reads 
\begin{equation}
\kappa(x)=\kappa_0 T(x)^{5/2}. 
\end{equation}
When collisionality drops, the classical Fourier law fails in describing heat transport, and the expression above leads to overestimated heat fluxes (see e.g \cite{Stangeby,Funda05} and references therein).\\ 
\indent Typically, in order to avoid unphysical divergences in the SH expression for the heat flux, an {\it ad hoc} flux limiter correction is introduced with the following harmonic average between the free streaming heat flux $q_{FS}=n v_{\rm th}T$ and the collisional expression $q_{SH}$:
\begin{equation}
q_{\parallel}=\left({\frac{1}{q_{SH}}}+{\frac{1}{\alpha n v_{\rm th}T}}\right)^{-1}.
\label{FL_expression}
\end{equation} 
In the formulae above $v_{\rm th}$ is the thermal velocity, $n$ the plasma density and $\alpha$ is a free parameter ranging from $0.1$ to $3$ characteristic values.\\
\indent In Figure \ref{Soledge_2D_temp} we report an example of the strong impact that such flux limiter expression can have on the predictions obtained from transport codes on energy fluxes at the wall. We consider a SOLEDGE2D simulation for WEST configuration \cite{JB14} in pure Deuterium with an input power $P_{in}=4MW$ and a gas puff activated in the private flux region with an injection rate equal to $4\times 10^{21}$ atoms per second. 
\begin{figure}
\begin{center}
\includegraphics[width=0.5\textwidth]{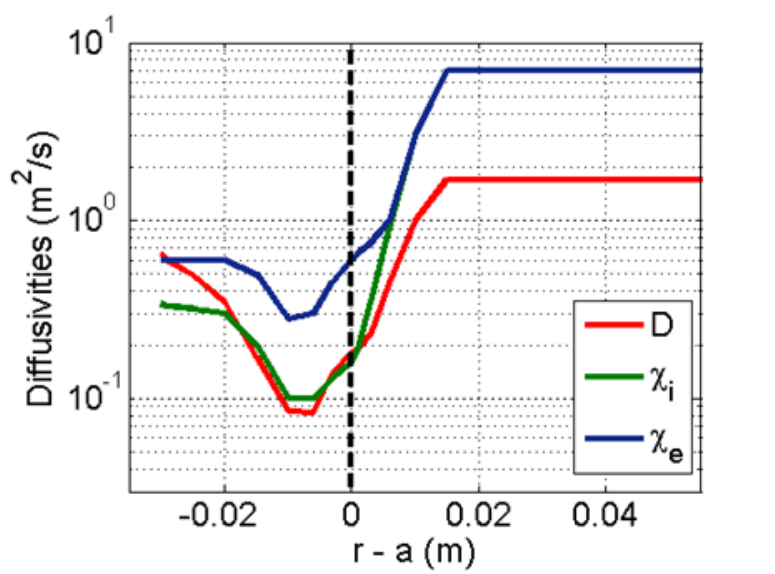}
\end{center}
\caption{Radial profiles of flux surface averaged transport coeffcients used in SOLEDGE2D-EIRENE simulations inspired from Ref.\cite{chankin}}
\label{rad_transp_coeff_S2D}
\end{figure}
The radial transport coefficients $D$ for density, $\chi_i$ for ion temperature and $\chi_e$ for electron temperature are reported in Fig.\ref{rad_transp_coeff_S2D}. They are settled equal to the ones presented in \cite{GC_NME17} and, waiting for measurements on WEST plasmas, have been chosen taking into account parameters which have been adjusted to match experimental mid-plane profiles of a H-mode ASDEX Upgrade plasma (see Ref. \cite{chankin}). 
The SOLEDGE-EIRENE simulations are performed considering three different expressions for the electron heat flux transport (while the ion heat flux is always computed using the flux limiter expression with $\alpha=0.2$): in the first simulation we do not activate the flux limiters (FL) and the electron heat flux is computed using the SH expression. In the second and third simulations the electron heat flux is computed using the flux limiter expression given by Eq.(\ref{FL_expression}) with $\alpha=0.3$ and $\alpha=0.15$, respectively. In Figure \ref{Soledge_2D_temp} (panel a), we show with a 2D colour map the electron temperature in a WEST poloidal section obtained in the SH case. The comparison between the electron temperature profiles at the outer strike point obtained from these three different cases (SH, FL with $\alpha=0.3$ and FL with $\alpha=0.15$) is presented in panel (b). We note that there is a strong reduction of the temperature peak value between the SH case and the one computed with a FL equals to $\alpha=0.15$.\\ 
\indent In order to improve the presently used ad hoc flux limiter treatment of parallel heat flux transport in edge plasma codes we consider a fluid description with the generalized version of the Fourier law implementing a non-local kernel for the heat flux computation as proposed, for example, in the paper by Luciani and co-authors \cite{Luciani83} 
\begin{equation}
  \label{eq:Luciani}
  q_{NL}(x)=- \int w(x,x^\prime) \kappa(x^\prime) \nabla T(x^\prime) d x^\prime,
\end{equation}
where $q_{NL}$ is the {\it non-local} heat flux, $w$ the delocalization kernel, $\kappa$ the classical Spitzer-H\"arm collisional conductivity and $T$ the temperature. The simplest phenomenological form of the kernel is the one in which memory decays exponentially in space,
\begin{equation}\label{eq:w}
w(x,x^\prime)=\frac{1}{2\lambda(x^\prime)} {\rm exp}\left(-|x-x^\prime|/\lambda(x^\prime)\right),
\end{equation}
where $\lambda(x^\prime)$ is the local electron mean free path at the position $x^\prime$.
\\
We have shown in Ref. \cite{HBCPP17} that solving the equation $\partial_x q_{NL}(x)=S$ using this non-local expression for the heat flux computation can lead to discontinuities in the temperature profile if the source term $S$ is very localized in space, such as for example, in the case of the interaction with the wall, and the collisionality takes medium and low values as it can happen in the scrape-off layer (SOL) plasma of tokamaks.
In order to overcome this issue we have introduced (see again Ref. \cite{HBCPP17}) the following expression for the heat flux:
\begin{equation}
\label{q_nonlocal}
  q_{NL,T}(x) = \widetilde{q_{NL} (x)} + q_{BC,0} exp\left(-\frac{x}{\lambda} \right) + q_{BC,L_\parallel} exp \left( \frac{x-L_\parallel}{\lambda} \right)
\end{equation}
This expression exhibits a first term describing the non-local heat flux computed from the continuous temperature
gradient expression in the plasma. The two last terms represent the impact of the boundary condition in the heat flux,
effect that decays exponentially away from the wall.They describe the long range influence of the boundary conditions.
The values $q_{bc,0}$ and $q_{bc,L_{\parallel}}$ are adjusted to match the sheath boundary condition for the heat flux, namely $q_{se}=\gamma n_{wall} c_s T_{wall}$ at both ends where $\gamma$ is the so called sheath transmission coefficient. We note that the value of the sheath heat transmission coefficient gamma depends also on the collisionality of the system and can take very large values when the high-energy tail exists (see for example Ref.\cite{tskhakaya08} and \cite{froese12}). However, for steady state condition like the one considered in this paper, the sheath transmission coefficients are quite constant for a large range of collisionality values.    
\section{Non-local heat transfer in fluid models: Application to 1D Scrape-off layer with localized particle and energy sources}\label{SOL1D_fluid}
We consider a 1D model of SOL plasma where we solve the standard equations for density, parallel momentum and ion and electron energy balance with standard Bohm boundary conditions, including the non-local expression for heat flux introduced above.
Localized sources of density (particle recycling) and energy (e.g. RF heating for both electrons and ions) have been added as follows.
For the particle source, simulating a recycling source term we have imposed
\begin{equation}
\label{particle_source1D}
S_n(x)=S_n^0 \left[ {\rm exp} \left(-\frac{x}{0.1L_\parallel} \right)+{\rm exp}\left( - \frac{L_\parallel - x}{0.1 L_\parallel} \right) + 0.005 \right],
\end{equation}
while for the energy sources, we have used Gaussian shaped sources located at the middle of the field line. The width of the energy source is controlled by $\lambda_E$ and reads
\begin{equation}
\label{energy_source1D}
  S_{Ee,i} =  S_{Ee,i}^0 {\rm exp} \left( - \left( \frac{x}{\lambda_E}-\frac{L_\parallel}{2 \lambda_E} \right)^2 \right).
\end{equation}
We report here two cases obtained varying the amplitude of the energy source and producing a first case at medium collisionality $\nu^{\star}=60$ and a second one at low collisionality with $\nu^{\star}=4$ where $\nu^{\star}=L_{\parallel}/\lambda$ with $\lambda$ the electron mean free path.
\begin{figure}
\begin{center}
\includegraphics[width=0.9\textwidth]{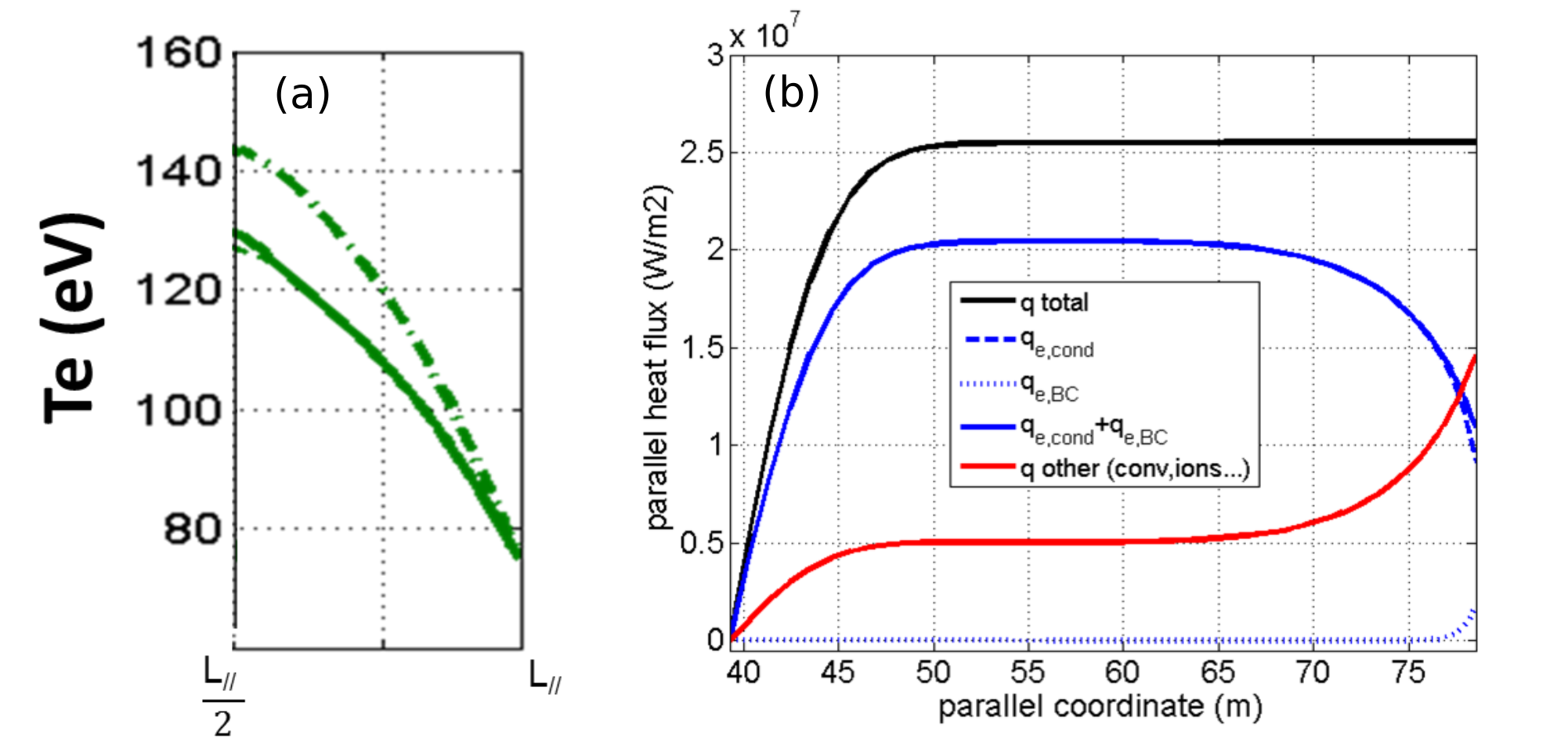}
\end{center}
\caption{Electron temperature profile along x parallel to magnetic field $B$ from upstream position ($x = 39 m$) to the wall ($x = 78 m$) obtained from SOLEDGE1D simulation at $\nu^{\star}=60$. The solid line represents the results obtained considering the non-local heat flux expression, the dashed line using the SH expression and the dashed-dotted line the flux limiter expression with the free parameter $\alpha=0.15$ (a). Heat flux profile along x parallel to $B$ from upstream position to the wall. The solid blue line is the sum of the contribution from the electron conductive term (reported with dashed line and dominant in this case) and 
the contribution from the $q_{e,BC}$ expression (see Eq. (\ref{q_nonlocal})) represented in dotted line, representing the long-range influence of the boundary conditions and very small in this case apart very closely to the target. The red solid line is the contribution from the remaining extra terms (e.g. convective, ions), while solid black line is the total heat flux (b).}
\label{Soledge1D_nustar60}
\end{figure}
We note that, for the medium collisionality case (see Fig. \ref{Soledge1D_nustar60}) the non-local expression collapses onto the standard SH expression with very small contribution coming from the non-local terms related to the influence of the boundary conditions. However, the $q_{e,BC}$ contribution is, as expected, non-negligible very close to the wall, see again the dotted line in Fig. \ref{Soledge1D_nustar60}.
\begin{figure}
\begin{center}
\includegraphics[width=0.9\textwidth]{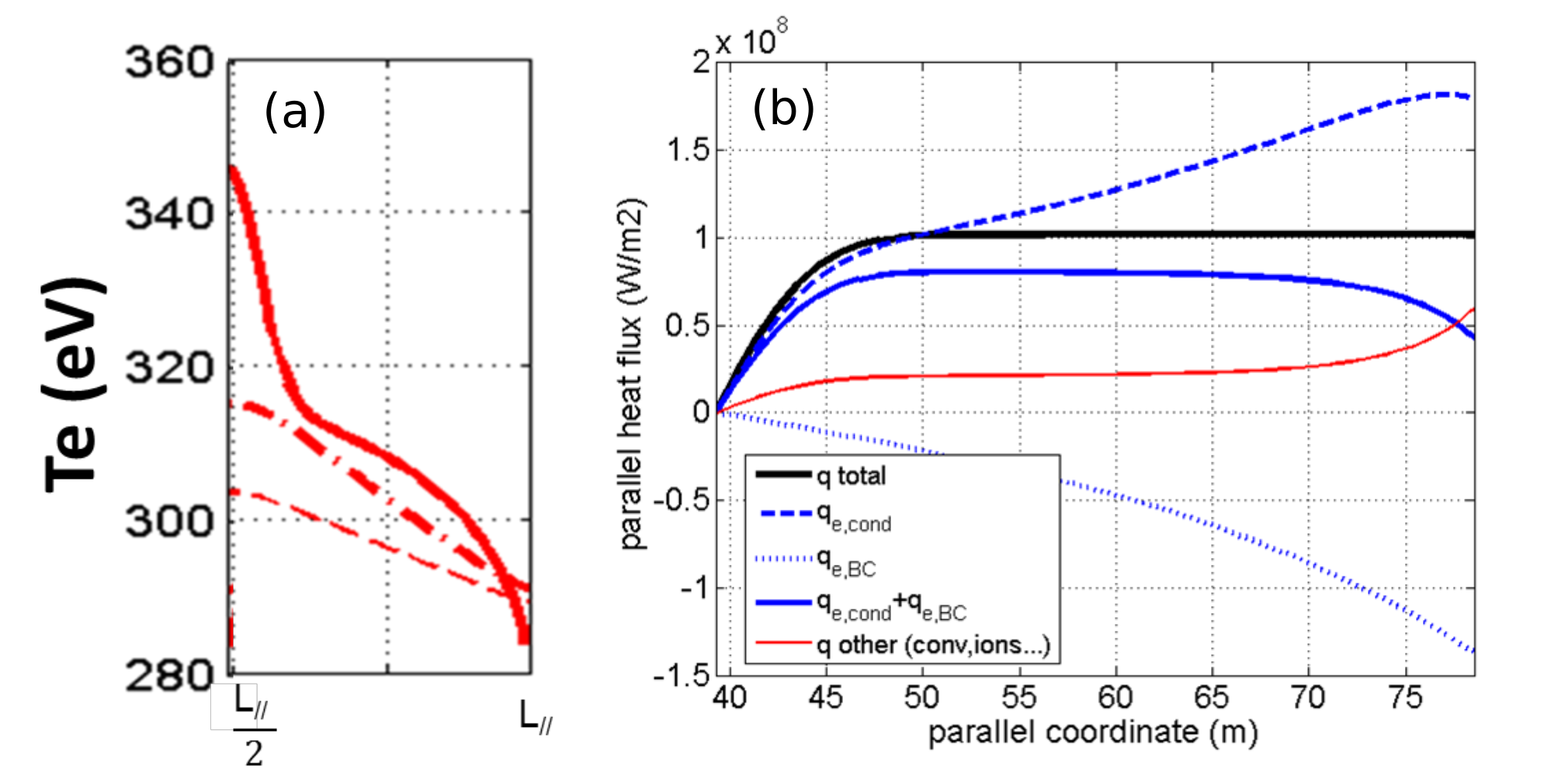}
\end{center}
\caption{Electron temperature profile along x parallel to magnetic field $B$ from upstream position ($x = 39 m$) to the wall ($x = 78 m$) obtained from SOLEDGE1D simulation at $\nu^{\star}=4$. Line types in figures are the same as those in Fig. \ref{Soledge1D_nustar60} (a). Heat flux profile along x parallel to $B$ from upstream position to the wall. Line types in figures are the same as those in Fig. \ref{Soledge1D_nustar60}. In this case the $q_{e,BC}$ contribution (dotted blue line) related to non-local effects is very large (b).}
\label{Soledge1D_nustar4}
\end{figure}
On the contrary, when the collisionality drops, the non-local expression is able to take into account the influence of the boundary conditions on the whole domain. In Fig.\ref{Soledge1D_nustar4} it appears clear that the contribution from the $q_{e,BC}$ expression to the total parallel heat flux is non-negligible on the entire domain and of the same order of magnitude of the conductive part. Interestingly, thanks to the proposed non-local expression, we can also recover the shape of the energy source into the temperature profile, which is Gaussian in the energy source region (see Fig. \ref{Soledge1D_nustar4} panel a).\\ 
\indent In next section we introduce the kinetic modeling of heat transfer which will be used for a first analysis of the results obtained in this section.
\section{Kinetic modelling of heat transfer}\label{model2}
From the kinetic point of view, weakly collisional plasmas are usually studied in terms of their phase-space distribution function $f(\mathbf{r},\mathbf{v})$ by means of the so-called Vlasov-Fokker-Planck equation (see e.g. \cite{vfp1,vfp2,vfp3}) that reads for the electron components as
\begin{equation}\label{vfp}
\frac{\partial f}{\partial t}+\mathbf{v}\cdot\nabla_\mathbf{r}f-\frac{e}{m_e}\left(\mathbf{E}+\frac{\mathbf{v}}{c}\times\mathbf{B}\right)\cdot\nabla_{\mathbf{v}}f=\nabla_{\mathbf{v}}\cdot(\nu\nabla_{\mathbf{v}}f).
\end{equation}
In the equation above, the diffusion coefficient $\nu$ appearing in the velocity-space diffusive term at the rhs could be in principle an explicit function of velocity, or be dependent on position through the local number density $n(\mathbf{r})=\int f{\rm d}\mathbf{v}$, see \cite{macdonald1957,rosenbluth1957}.\\
\indent Equation (\ref{vfp}) can be easily integrated with standard implicit Eulerian codes in the 1D1V and 1D2V cases \cite{vfp2}, adopting standard Maxwell solvers to account for the self-consistent electric and magnetic fields $\mathbf{E}$ and $\mathbf{B}$. Already in 2 spatial dimensions (2D2V or 2D3V) such approach rapidly becomes numerically expensive and we therefore rely on particle\footnote{Note that particles are to be thought as a discrete sampling of $f$, rather than actual ``particles".} based (semi-)Lagrangian methods such as particle-in-cell (PIC, see e.g. \cite{compsim}). Including the contribution of collisions in cell-based PIC codes is usually time consuming and model dependent, here we used a stochastic approach based on the multi-particle collision (hereafter MPC) technique.  
\subsection{Multi-particle collision method}\label{MPC}
Originally introduced by Malevanets and Kapral \cite{1999JChPh.110.8605M} for the simulation of complex fluids (e.g. polymers in solution, colloidal fluids), in 3 spatial dimension, the MPC scheme partitions the system of $N_p$ particles in $N_c$ cells\footnote{In our implementation the mesh used in the MPC step is the same as the one used by the PIC code to compute electromagnetic fields.}. Between two standard propagation steps,
inside each cell the particle velocities in the cell's centre of mass $\delta\mathbf{v}_j=\mathbf{v}_j-\mathbf{u}_i$ are rotated of an angle $\varphi$ around a random axis $\mathbf{R}$ and then converted back to the simulation frame, so that for the $j-$th particle in cell $i$ 
\begin{equation}\label{rotation}
\mathbf{v}_{j}^\prime=\mathbf{u}_i+\delta\mathbf{v}_{j,\perp}{\rm cos}(\varphi)+(\delta\mathbf{v}_{j,\perp}\times\mathbf{R}){\rm sin}(\varphi)+\delta\mathbf{v}_{j,\parallel},
\end{equation}  
where $\delta\mathbf{v}_{j,\perp}$ and $\delta\mathbf{v}_{j,\parallel}$ are the relative velocity components perpendicular and parallel to $\mathbf{R}$, respectively. Such operation exactly conserves in each cell the total kinetic energy $K_i$ and the three components of the momentum $\mathbf{P}_i$. For an extensive proof of the conservation laws see Appendix A in\cite{2017PhRvE..95d3203D}. In addition, it is also possible to conserve the component of the angular momentum $\mathbf{L}$ parallel to $\mathbf{R}$ by choosing $\varphi$ so that 
\begin{equation}\label{sincos}
{\rm sin}(\varphi)=-\frac{2a_ib_i}{a_i^2+b_i^2};\quad  {\rm cos}(\varphi)=\frac{a_i^2-b_i^2}{a_i^2+b_i^2},
\end{equation}
with cell-dependent coefficients $a_i$ and $b_i$ given by
\begin{equation}\label{ab}
a_i=\sum_{j=1}^{N_i}\left[\mathbf{r}_j\times(\mathbf{v}_j-\mathbf{u}_i)\right]|_z;\quad b_i=\sum_{j=1}^{N_i}\mathbf{r}_j\cdot(\mathbf{v}_j-\mathbf{u}_i).
\end{equation}
In the formulae above, $\mathbf{r}_j$ are the particles position vectors, and the notation $|_z$ means that one is taking (without loss of generality) the component of the vector $\mathbf{A}_i$ parallel to the $z$ axis of the simulation's coordinate system.\\
\indent For two dimensional systems, Equation (\ref{rotation}) reduces to $\mathbf{v}_{j}^\prime=\mathbf{u}_i+\mathbf{G}_{\varphi,i}\cdot\delta\mathbf{v}_{j}$,  
where $\mathbf{G}_{\varphi,i}$ is a 2D rotation matrix of an angle $\varphi$ chosen according to relations (\ref{sincos},\ref{ab}), see Ref.\cite{2017PhRvE..95d3203D}. In both 2D and 3D cases, the generalization to multi-mass models is straightforward and implies the substitution of velocity vectors with momentum vectors.\\
\indent In one dimension, the multi-particle collision involves instead a velocity sign inversion with a momentum shift (see also \cite{2015PhRvE..92f2108D}) and the two conserved quantities are the linear momentum $P_i$ and the kinetic energy $K_i$. During the collision step the stochastic momentum shifts $w_j$ are extracted for each particle from a normal distribution depending on the cell temperature, so that the conservation of $P_i$ and $K_i$ now reads
\begin{eqnarray}\label{sist}
P_i&=&\sum_{j=1}^{N_i} m_jv_{j}=\sum_{j=1}^{N_i} m_jv^\prime_{j}=\sum_{j=1}^{N_i} (c_iw_j+d_im_j);\nonumber\\
K_i&=&\frac{1}{2}\sum_{j=1}^{N_i} m_jv_{j}^2=\frac{1}{2}\sum_{j=1}^{N_i} m_jv_{j}^{\prime 2}=\frac{1}{2}\sum_{j=1}^{N_i} m_j(c_iw_j/m_j+d_i)^2,
\end{eqnarray}
where $N_i$ is the number of particles in cell $i$, $m_j$ and $v_j$ are the $j$-th particles mass and velocity, and $c_i$ and $d_i$ are unknown cell-dependent quantities. Eqs. (\ref{sist}) constitute a linear system that to be solved for $c_i$ and $d_i$. 
We define the stochastic momentum and kinetic energy increments 
\begin{eqnarray}
P_i^*=\sum_{j=1}^{N_i} w_{j}; \quad K_i^*=\frac{1}{2}\sum_{j=1}^{N_i} {w_{j}^2}/{m_j},
\end{eqnarray}
and rescale them, together with $P_i$ and $E_i$, by the total mass in cell $i$,  $M_i=\sum_{j=1}^{N_i} m_j$ as  
$\tilde{P_i^*}=P_i^*/M_i$, $\tilde{P_i}=P_i/M_i$, $\tilde{K_i^*}=K_i^*/M_i$ and $\tilde{K_i}=K_i/M_i$. The coefficients $c_i$ and $d_i$ are then easily computed as
\begin{equation}
c_i=\sqrt{\frac{{2\tilde K_i-\tilde P_i^2}}{{2\tilde{K_i^*}-\tilde{P_i^{*2}}}}};\quad d_i=\tilde P_i-\tilde{P_i^*}c_i,
\end{equation}
so that the new velocities after the multi-particle collision finally read $v^\prime_{j}=c_iw_j/m_j+d_i$.\\
\indent In a series of papers on the anomalous diffusion and heat transfer in 1D one-component plasmas \cite{2010JPhCS.260a2005B,2013PhRvE..87b3102B,2015PhRvE..92f2108D,2017PhRvE..95d3203D}, we have applied a hybrid PIC-MPC technique where velocity exchange inside the cells is conditioned to an interaction probability $\mathcal{P}_i$ dependent on the local plasma parameters, in order to account for Coulomb collisions in a more physical way and also to treat spatially and thermally inhomogeneous systems.\\    
\indent In each cell we define the species-averaged {\it plasma coupling parameter}
\begin{equation}
\bar\Gamma_i=\frac{E_{C,i}}{k_BT_i},
\end{equation}
where $E_{C,i}=\langle q^2\rangle_i/4\pi\epsilon_0\xi_i$ is the mean Coulomb energy per particle, $\langle q^2\rangle_i$ the particles average (squared) charge in cell $i$, and $\xi_i$ is a typical inter-particle distance depending on the local particle number density $n_i$, finally, the cell temperature $T_i$ is assumed to be proportional to the average kinetic energy of the particles inside the cell as $k_BT_i=(1/N_i)\sum m_jv_j$.
Before the collision step, the code evaluates for each cell the (multi-particle) collision probability as 
\begin{equation}\label{prob}
 \mathcal{P}_i=\frac{1}{1+\bar\Gamma_i^{-2}}.
\end{equation}
After sampling a random number $\mathcal{P}_i^*$ from a uniform distribution in the interval $[0,1]$, the multi-particle collision happens if $\mathcal{P}_i^*/\mathcal{P}_i\leq1$.
\subsection{Preliminary 1D kinetic simulations}
Here we present numerical simulations of 1D systems modelling the plasma dynamics along a field line between a hot thermal bath (upstream region) and the colder wall region. In this preliminary work we always assume regimes of strong correlation between ion and electron motion as well as fulfillment of quasi-neutrality condition. In such conditions, the main contribution to the heat flux is due to electrons (see panels (b) of Figs. \ref{Soledge1D_nustar60} and \ref{Soledge1D_nustar4}), we therefore consider a single component system  representing the electrons and treat the ions as a non-interacting background adjusting itself as the electron density $n_e$ evolves, in order to yield a globally null electric field. With such assumptions Equation (\ref{vfp}) becomes a standard one dimensional Fokker-Planck equation of the form $\partial_t f +v\partial_r f=\partial_v(\nu\partial_v f)$.\\
\indent In our PIC-MPC code the interaction with the hot source and the wall is modeled with standard Maxwellian thermal baths. In practice, when a simulation particle enters the hot region its velocity $v$ is substituted with a new velocity $v^\prime$ taken from a Maxwellian distribution at temperature $T_{\rm Hot}$.
when instead the particle hits the cold wall, it is either reflected elastically, or re-immitted in the simulation domain with a velocity taken from a Maxwellian distribution at temperature $T_{\rm Cold}$, with probabilities one-half. Note that, with such choice, the total particle number $N_p$ is conserved as no particle leaves the system. In principle, it is also possible to account for particle evaporation by considering an additional velocity-dependent exclusion protocol that selects hotter particles and removes them from the system. A ''stochastic evaporation" algorithm is currently under testing and will be discussed in a forthcoming publication.\\
\begin{figure}
\begin{center}
\includegraphics[width=0.9\textwidth]{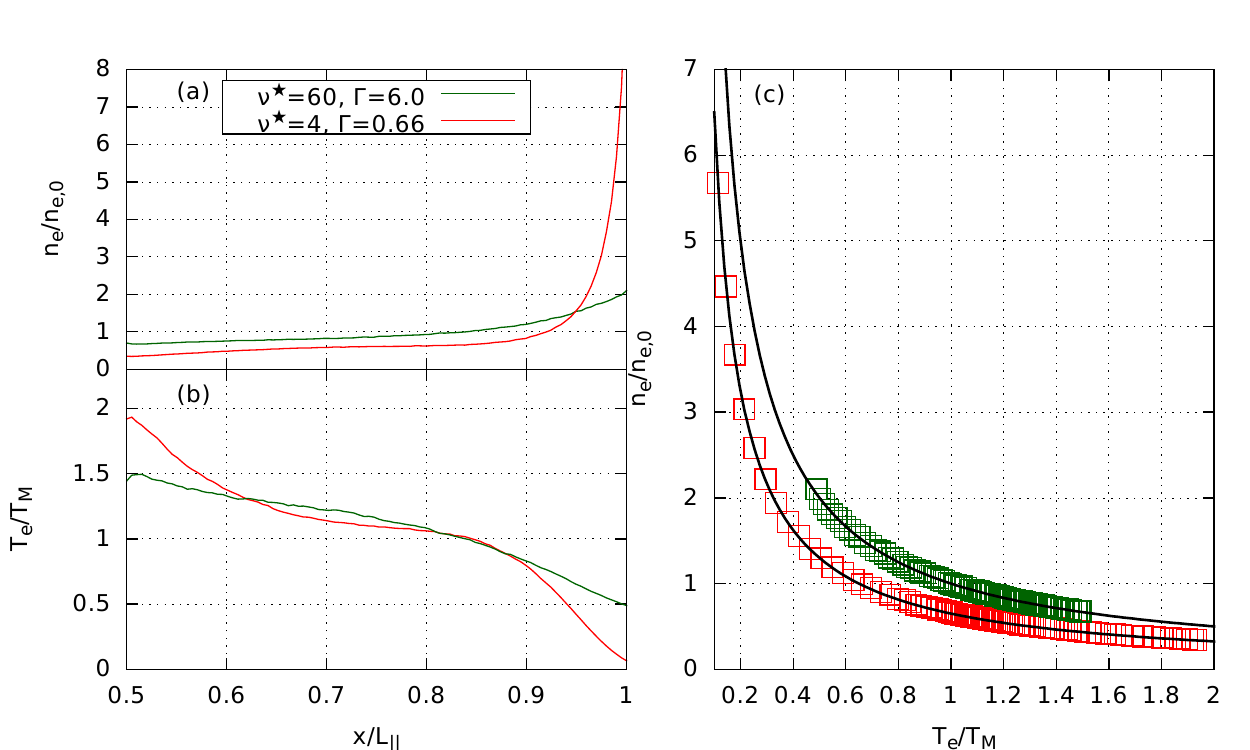}
\end{center}
\caption{For two models with $\Gamma=6$ and 0.66: final electron number density $n_{e}$ (a) and temperature profile $T_e$ (b) as function of the parallel coordinate $x/L_{||}$, and density temperature relation (squares) and best-fit curves (solid lines), (c). Parallel density and temperature profiles are given only for $L_{||}/2\leq x\leq L_{||}$, as they are perfectly symmetrical in the other half of the simulation domain.}
\label{rhotemp}
\end{figure}
\indent Particle propagation is carried out with a standard second order leap-frog scheme while collisions are accounted for as described in Sec. \ref{MPC}. All simulations discussed here were performed with fixed timestep $\delta\tau=0.01\Omega_P^{-1}$, where $\Omega_P=\sqrt{n_ee^2/m_e\epsilon_0}$ is the plasma frequency of the system neglecting the thermal motion, and extended up to $\tau=10^3/\Omega_P$.\\
\indent In the kinetic simulations we have taken the same combinations of temperature, density and parallel length as in the two cases discussed in Sect. \ref{SOL1D_fluid}, yielding the two values of the collisionality $\nu^\star=60$, 4. We have assumed equilibrium initial conditions by placing the particles representing the electron component homogeneously on the simulation domain $\left[0, L_{||}\right]$ (i.e. constant initial electron number density $n_{e,0}$), with velocities taken from a thermal distribution at temperature $T_{e,0}$. After a short transient of about $10\delta\tau$ the thermal baths at $T_{\rm hot}=130$ eV and $T_{\rm cold}=78$, and $T_{\rm hot}=345$ eV and $T_{\rm cold}=285$ eV for the $\nu^\star=60$ and $\nu^\star=4$ cases are applied for both cases in $x=L_{||}/2$ and $x=0$, $L_{||}$.\\
\indent From the initial values of the electron temperature and density $T_{e,0}$ and $n_{e,0}$  we derive the initial global plasma coupling parameter $\Gamma=E_{C,0}/k_BT_0$ that gives another measure of how strong is the system's collisionality (at least) in its initial state (i.e., at fixed $L_{||}$ or at fixed $\Omega_P$, larger $\Gamma$ implies higher collisionality). With the present  combination 
\begin{figure}
\begin{center}
\includegraphics[width=0.9\textwidth]{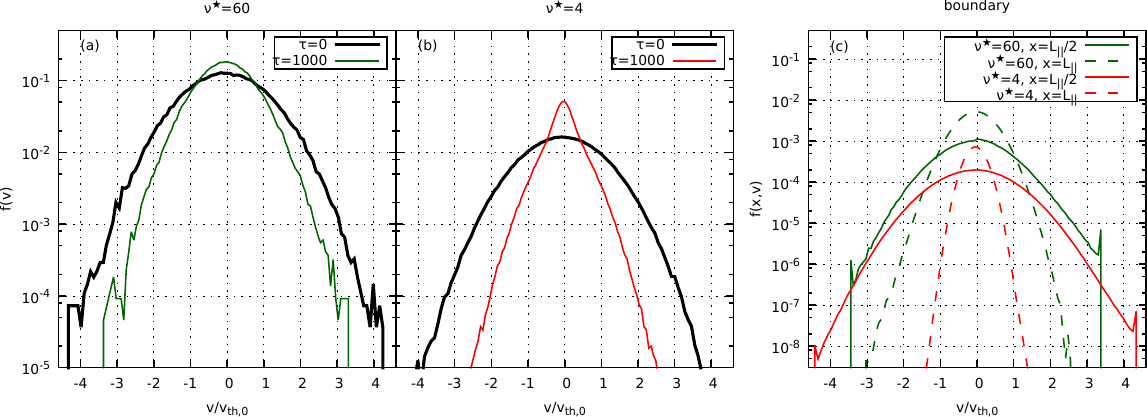}
\end{center}
\caption{Velocity distributions $f(v)$ for the case with $\nu^\star=60$ (a) and $\nu^\star=4$ (b). The thin solid lines correspond to the equilibrium state (reached at around $\tau\approx 10^3$), while the heavy solid line mark the initial velocity distributions. Sections of the phase-space distribution function $f(x,v)$ at the cold wall (dashed lines) and hot (solid lines lines) (c). In all cases the velocities are normalized with respect to the initial thermal velocity $v_{{\rm th},0}$.}
\label{frv}
\end{figure}
of parameters we obtain $\Gamma\approx 6$ for $\nu^\star=60$ and $\Gamma\approx 0.66$ for $\nu^\star=4$.\\
\indent Figure \ref{rhotemp} shows the asymptotic equilibrium state of the two models with collisionality $\nu^\star=60$ and 4 ($\Gamma=6$ and 0.66), in contact with a hot source at $x=L_{||}/2$ and a cold wall at $x=0$ and $L_{||}$. In both cases the systems (consistently) show non-uniform density profiles with a density accumulation in correspondence of the cold source. Vice-versa, the density depletes approaching the hot source, due to the larger mean velocities of particle in this region. At variance with Figs. \ref{Soledge1D_nustar60}-\ref{Soledge1D_nustar4}, the electron temperature $T_e$ is given in units of the system's mean final temperature $T_M$, so that the two curves can be more easily compared being on the same scale. The initially more collisional system (i.e. $\Gamma=6$) has a quasi-linear temperature profile over a broader interval of the parallel coordinate $x$ (i.e. $0.5\leq x/L_{||}\leq 0.73$), while the weakly collisional system has a more complex asymptotic temperature profile characterized by several slope changes and a flat central region (remarkably similar to the corresponding curve in Fig. \ref{Soledge1D_nustar4}, panel a), pointing to a highly non-local heat transport regime. Remarkably, in both cases the final electron pressure $P_e\propto n_eT_e$ is spatially constant as we clearly observe $n_e\propto T_e^{-1}$ at $\tau=10^3$ (panel c).\\
\indent Figure \ref{frv} shows for the same systems of Fig. \ref{rhotemp} the initial and final velocity distributions $f(v)$ (panels a, b) and the sections of (half of) the numerically-recovered phase-space distribution function $f(x,v)$ at $x=L_{||}/2$ and $x=L_{||}$ (panel c). The strongly interacting model with $\nu^\star=60$ presents a final $f(v)$ that is well described by a Gaussian, while the model with $\nu^\star=4$ has a clearly non-thermal final velocity distribution. Both cases, however, appear to be colder in their final state with respect to their initial states. For what concerns the phase-space distribution, while in both cases $f(x,v)$ clearly approaches a thermal distribution in correspondence of the cold point, the structure of $f(x,v)$ at $x=L_{||}/2$ is somewhat more complicated and characterized by a fatter tail at positive velocities (i.e. corresponding to particles moving {\it towards} the cold point). In addition, in correspondence of the highest velocities attained by the particles, two peak-like structures can be clearly seen. We interpret this feature as a finite-size effect due to the almost vanishing life-time of larger velocities reaching the cold wall. In fact, at fixed $\nu^\star$, $\Gamma$, and thermal baths temperatures $T_{\rm hot}$ and $T_{\rm cold}$, such peaks tend to disappear for increasing $L_{||}$.     
\section{Conclusion and outlook}
We have shown the impact of flux limiter techniques on the computation of heat flux on divertor tokamak simulations. We have proposed the implementation of a non-local approach in a 1D fluid model and we have rpesented the numerical results obtained with SOLED1D at medium and high collisionality. In the second part of the paper a PIC-MPC kinetic simulations are presented. they offer a particle-based approach that appears to be more suitable to study transient regimes and relaxation processes. Remarkably, for the case studies discussed in this paper, we found good agreement between this approach and the fluid modelling, suggesting that further evolutions of the fluid scheme could be tested against more detailed particle-in-cell-MPC simulations including more species and the effect of the self-consistent fields.    
\section*{Acknowledgments}
This work has been carried out within the framework of the EUROfusion Consortium and has received funding from the Euratom research and training program 2014-2018 under grant agreement No 633053 for the project WP17-ENR-CEA-01. The views and opinions expressed herein do not necessarily reflect those of the European Commission. This work was granted access to the HPC resources of Aix-Marseille
University financed by the project Equip@Meso (ANR-10-EQPX-29-01) of
the program ``Investissments d'Avenir'' supervised by the Agence
Nationale pour la Recherche. P.F.D.C. acknowledges partial support by the INFN project DYNSYSMATH 2017.

\end{document}